\documentclass[journal]{IEEEtran}
\usepackage{adjustbox}
\usepackage{courier} 
\usepackage{listings, xcolor}
\usepackage{cite}
\usepackage{amsmath,amssymb,amsfonts}
\usepackage{algorithmic}
\usepackage{graphicx}
\usepackage{textcomp}
\usepackage{xcolor}

\usepackage{todonotes}
\usepackage{hyperref}
\usepackage{multirow}
\usepackage{lmodern,babel,adjustbox,booktabs}
\usepackage{subcaption}
\usepackage{diagbox}
\usepackage[T1]{fontenc}
\usepackage[most]{tcolorbox}
\usepackage{lmodern} 
\usepackage{lipsum}
\usepackage{url}
\usepackage{lmodern,babel,adjustbox,booktabs,multirow}
\usepackage{tabulary}
\usepackage[utf8]{inputenc}

\usepackage{enumitem}
\setlist[itemize]{noitemsep, topsep=0pt}

\usepackage{tcolorbox}
\usepackage[flushleft]{threeparttable}
\usepackage{fancyhdr}
\usepackage{lastpage}
\newtcolorbox{prbox}[3][]
{
  colframe = #2!25,
  colback  = #2!10,
  coltitle = #2!20!black,  
  title    = {#3},
  #1,
}

\begin{document}

\title{How Do Developers Structure Unit Test Cases? An Empirical Study from the  ``AAA''  Perspective
}


\author{Chenhao~Wei,~\IEEEmembership{Member,~IEEE,}
        Lu~Xiao,~\IEEEmembership{Member,~IEEE,}
        Tingting~Yu,~\IEEEmembership{Member,~IEEE,}
        Sunny~Wong,~\IEEEmembership{Member,~IEEE,}
        Abigail~Clune
\thanks{C. Wei and L. Xiao are with the School of Systems and Enterprises, Stevens Institute of Technology, Hoboken, NJ, 07030 USA. e-mails: cwei7@stevens.edu, lxiao6@stevens.edu.}
\thanks{T. Yu is with the Computer Science and Engineering Department at University of Connecticut, Storrs, CT 06269-4155, USA. e-mail: tingting.yu@uconn.edu}
\thanks{S. Wong is with Envestnet, Inc., Berwyn, PA 19312, USA. e-mail: sunny@computer.org}
\thanks{A. Clune is with AGI, an Ansys Company, Exton, PA 19341, USA. e-mail: abigail.clune@ansys.com}
}
{}




\maketitle
\begin{center}
\footnotesize
\textbf{©2025 IEEE.} This is the authors’ version of the manuscript accepted 
for publication in the \textit{IEEE Transactions on Software Engineering (TSE)}.\\
\textbf{Important note:} The official TSE version contains additional 
data and more comprehensive results.\\
Personal use of this material is permitted. Permission from IEEE must be obtained 
for all other uses.\\
\textbf{DOI of the published version:}
\href{https://doi.org/10.1109/TSE.2025.3537337}{10.1109/TSE.2025.3537337}
\end{center}
\begin{abstract}
The \textit{AAA} pattern, i.e. \textit{arrange}, \textit{act}, and \textit{assert}, provides a unified structure for unit test cases, which benefits comprehension and maintenance. However, there is little understanding regarding whether and how common real-life developers structure unit test cases following \textit{AAA} in practice. In particular, are there recurring anti-patterns that deviate from the \textit{AAA} structure and merit refactoring? And, if test cases follow the \textit{AAA} structure, could they contain design flaws in the \textit{A} blocks? If we propose refactoring to fix the design of test cases following the \textit{AAA}, how do developers receive the proposals? Do they favor refactoring? If not, what are their considerations?

This study presents an empirical study on 435 real-life unit test cases randomly selected from four open-source projects. Overall, the majority (71.5\%) of test cases follow the \textit{AAA} structure. And, we observed three recurring anti-patterns that deviate from the \textit{AAA} structure, as well as four design flaws that may reside inside of the \textit{A} blocks. Each issue type has its drawbacks and merits corresponding refactoring resolutions. We sent a total of 18 refactoring proposals as issue tickets for fixing these problems. We received 78\% positive feedback favoring the refactoring. From the rejections, we learned that return-on-investment is a key consideration for developers. The findings provide insights for practitioners to structure unit test cases with \textit{AAA} in mind, and for researchers to develop related techniques for enforcing \textit{AAA} in test cases.
\end{abstract}




\section{Introduction}

The \textit{AAA} pattern refers to the three section layout of writing a unit test case: \textit{arrange}, \textit{act}, and \textit{assert}~\cite{khorikov2020unit}. The \textit{AAA} pattern provides a natural and intuitive flow for creating a unit test case. In \textit{arrange}, the required environment, such as object creation and mock setup, is prepared. In \textit{act}, the target function being tested is executed. In \textit{assert}, the actual output from the \textit{act} is checked against expectation. A test failure is raised for attention when the actual output does not match the expectation. Following is an example test case that follows the \textit{AAA} structure. 

\begin{lstlisting}[language = Java, frame = single , numbers=right, firstnumber = 1 , escapeinside={(*@}{@*)}]
@Test
public void testGetByPrefix_Drop(){
    Config con = new Config();//arrange
    tc.set(PROP_PREFIX);//arrange
    var p = tc.getAllProperties();//act
    assertEquals("prop", p);}//assert
\end{lstlisting}

Although scientific evidence of its benefits from formal studies is absent, the \textit{AAA} pattern benefits the comprehension and maintenance of test cases, as being well advocated in technical blogs and tutorials. For example, ``\textit{Following this pattern does make the code quite well structured and easy to understand}''~\cite{blog1}; ``\textit{The AAA pattern is simple and provides a uniform structure for all tests in the suite. This uniform structure is one of its biggest advantages: once you get used to this pattern, you can read and understand the tests more easily. That, in turn, reduces the maintenance cost for your entire test suite}'';~\cite{blog2} and ``\textit{It is a structure or a way of thinking about and arranging your tests so that they can be clearly understood}''~\cite{blog3}. 

Despite the benefits, \textit{AAA} is only a structure, a way of thinking, or a guideline for writing test cases. There is no tool to enforce the \textit{AAA} pattern. Instead, the adoption and implementation of the \textit{AAA} pattern defer to the developers who actually write the test cases. A test case may not follow the \textit{AAA} pattern due to special consideration. For example, in the scenario of test-driven development~\cite{gulati2017java}, the test cases are written before the production functions are completed. The developer may start a test case with \textit{assert}, which describes the expected behavior. The test case is intended to fail with only the \textit{assert} when not enough understanding of the actual function is available. 

Developers may also imprudently violate the  \textit{AAA} pattern due to insufficient design. For example, a test case may contain multiple blocks of the \textit{AAA} structure, with each block \textit{arranges}, \textit{acts}, and \textit{asserts} for a different test scenario of the same function. Following is an example based on a real-life test case we observe. It combines different testing scenarios of \textit{getByPrefix} in one test case.
\begin{lstlisting}[language = Java, frame = single , numbers=left, firstnumber = 1 , escapeinside={(*@}{@*)}]
@Test
public void testGetByPrefix(){
    Config con = new Config();//arrange (*@\label{b1-start}@*)
    tc.set(PROP_PREFIX);//arrange
    var p = tc.getAllProperties();//act
    assertEquals("prop", p);//assert (*@\label{b1-end}@*)
    
    tc.set(SCAN_PREFIX);//arrange (*@\label{b2-start}@*)
    p = tc.getAllProperties();//act
    assertEquals("scan", p);}//assert(*@\label{b2-end}@*)
\end{lstlisting}
This is also a blunt violation of the ``\textit{single responsibility}'' principle in software design~\cite{martin2003agile, masel1996principles}. The drawback is that the test case could fail due to either of the two scenarios, adding extra difficulty to comprehension, maintenance, and debugging. This test case with multiple \textit{AAA} blocks should be broken down into separate test cases with each focusing on one test scenario and failing only due to that one scenario, to advocate the \textit{AAA} design and the ``\textit{single responsibility}'' principle.


In the existing literature, there is little understanding of whether and to what extent real-life test cases actually follow the \textit{AAA} pattern. Is this pattern only a theory, or is it widely practiced? In particular, in test cases that do not follow the \textit{AAA} pattern, are there recurring anti-patterns that deviate from the \textit{AAA} design and merit from refactoring? And, if test cases follow the \textit{AAA} pattern, could they still contain design flaws in the \textit{A} blocks that merit attention? If we propose refactoring to these test cases, how do the developers receive the proposals? Do they favor an investment on refactoring for enforcing the \textit{AAA} pattern? If not, what are their considerations? These are the questions that we are interested in addressing in this study. This fills the gap in the empirical knowledge regarding the practice of the \textit{AAA} pattern in test cases. The objective is to provide insights for practitioners in creating test cases with \textit{AAA} in mind, and for researchers in developing facilitating techniques.

This work is highly related to research in test smells, which focus on surface indications of deeper problems in test code, according to Fowler~\cite{beck1999bad}. However, our work distinguishes itself in two ways: 1) It focuses on root-cause revealing design flaws and anti-patterns by leveraging the holistic \textit{AAA} context in a test case. In comparison, test smells usually stay at the surface of problem indications. And, 2) it reveals four new design issues in test cases that have not been reported in prior work. The detailed comparison with test smells is resented in Section~\ref{sec:related}.

This work makes the following contributions:
\begin{itemize}
    \item First of its kind empirical study to investigate whether and how often the \textit{AAA} pattern is practiced.
    \item Novel design flaws and anti-patterns in test cases, reasoned based on the holistic context of \textit{AAA} in test cases, which could shed light on design root causes to surface problem indications. 
    \item Real-life developers' perspectives and considerations regarding whether they favor fixing the design problems under the \textit{AAA} context.
\end{itemize}

\section{Research Questions}\label{sec:rqs}

\textit{\textbf{RQ1: How often do real-life test cases follow the \textit{AAA} pattern? How do the \textit{AAA} test cases and \textit{anti-AAA} test cases compare to each other?}} Although \textit{AAA} is advocated by textbooks, tutorials, and blogs, it is unclear how often real-life test cases actually follow the \textit{AAA} pattern. We aim to report the percentage of real-life test cases that actually follow this pattern. In addition, we are also interested to compare test cases that follow \textit{AAA} with those that violate \textit{AAA} pattern, in terms of their general complexity measured by the LOC and Cyclomatic metrics, as well as their layout structure measured by the number of \textit{arrange}, \textit{act}, and \textit{assert} statements. This helps us to understand what is the key difference between \textit{AAA} cases and the cases that are not \textit{AAA}.
    
    
\textit{\textbf{RQ2: What are common ways that test cases deviate from the \textit{AAA} pattern? What are some anti-patterns that can occur from this deviation? In addition, do \textit{AAA} test cases contain design flaws that merit improvement?}} To answer this RQ, we manually inspect each test that does not follow the \textit{AAA} pattern. The goal is to reveal in what ways the \textit{AAA} pattern is not followed. And based on the observations, we summarize recurring violations of the \textit{AAA} pattern, i.e. the anti-patterns. For investigating the design flaws in \textit{AAA} cases, we specifically investigate design features related to control flow, such as the usage of \textit{if-else}, \textit{try-catch}, and \textit{while/for} loops, which could add complexity to the test cases. 
    
\textit{\textbf{RQ3: How do real-life developers receive the refactoring proposals to improve the \textit{anti-AAA} test cases and the \textit{AAA} cases with design flaws?}} We perform manual refactoring to restructure the \textit{anti-AAA} test cases so that after the refactoring, the \textit{AAA} pattern will be followed. We also perform refactoring to fix design flaws with test cases that already follow the \textit{AAA} structure to further improve them. We send issue reports with our tentative refactoring solution and see how real-life developers receive the refactoring proposals. 

\section{Approach}\label{sec:approach}

\subsection{Study Subjects}
  
\begin{table}[h]
    \centering
    \caption{Study Subjects}
    \label{tbl:subjects}
     \resizebox{\columnwidth}{!}{
    \begin{tabular}{|c|c|c|c|c|c|c|c|}
    \hline
    \multirow{2}{*}{Project Name} &
    \multirow{2}{*}{Version} &
    \multirow{2}{*}{\#Commits} &
    \multirow{2}{*}{\#Contributors} &
    \multicolumn{2}{|c|}{\#Files} &
    \multicolumn{2}{|c|}{\#Test Cases} \\ \cline{5-8}
    &&&&All&Test&All&Selected  \\
    \hline
    Accumulo&2.0.0&10,087&146&2,454&607&2,333&77\\
    \hline
    Druid&0.19.0&10,471&507&7324&1,297&6,532&239   \\
    \hline
    Cloudstack&4.13.1.0&32,250&366&7,816&514&3,002&96\\
    \hline
    Dubbo&2.7.7&4,307&424&6,524&1,376&2,765&88 \\
    \hline
    \end{tabular}}
\end{table}
Our study is based on four active, real-life open source projects, including Accumulo~\cite{accumulo}, Druid~\cite{druid}, Cloudstack~\cite{cloudstack}, and Dubbo~\cite{dubbo}. 
From each project, we randomly select about 3\% test cases from each project for our study, since we cannot afford to study all test cases in these projects. A total of 500 test cases are in our initial study dataset. Table~\ref{tbl:subjects} lists some basic facts about these projects, including the project name, studied version, the total number of contributors, the total number of commits, the number of source files and test files, as well as the total number of test cases and the number of selected test cases from each project. We study these projects because of the following rationale. First, these projects are in different domains. Second, these projects have a non-trivial amount of test code---514 to 3,794 test files---indicating that testing is of importance. Second, these projects are still actively developing and updating---with up to 507 contributors and up to 32,250 commits. Being active is important since we aim to collect feedback from their developers. 
The data that support the findings of this study are openly available in figshare via the link at \url{https://figshare.com/s/b1d6b70e10837aaf3f17}.

\begin{figure*}[h]
	\centering
    \includegraphics[width=\textwidth]{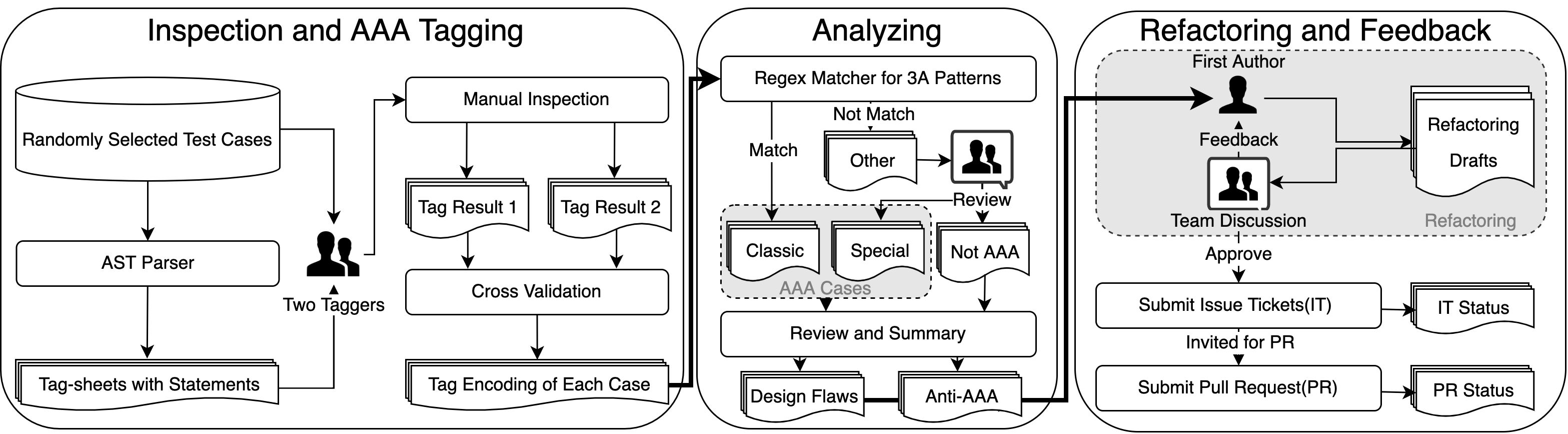}
	\caption{Experiment Approach}
	\label{fig:approach}
\end{figure*}

\subsection{Step 1: Test Case Inspection and \textit{AAA} Tagging}
In this step, we manually inspect each test case to tag the statements as \textit{Arrange}, \textit{Act}, or \textit{Assert}, based on a good understanding of the intention of the test case. 

\textbf{\textit{Taggers.}} To avoid personal bias, two taggers work on this task independently from each other. The first tagger is a Ph.D. student whose research is in Software Engineering. He has 2 years of prior working experience as a software developer. The second tagger is a master's student majoring in Computer Science. 

\textbf{\textit{``Tag-sheets.''}} To smooth the inspection and tagging process, we create a parser, which takes a test case as input, and outputs a full and expanded list of invocation statements from the test case. The parser is based on the ``abstract syntax tree (AST)'' analysis of each test case. Each method invocation in the test case is expanded into its internal call trace until cannot be further expanded. Note that the parser does not expand the invocation trace from the production methods, since a production method should be tagged ``atomically'' as one of the ``A''s---usually \textit{arrange} or \textit{act}. The output from the parser serves as our ``tag-sheet'' for each test case, which is used together with the code base in the inspection and tagging process. The taggers use the ``tag-sheet'' to annotate the type of each statement in the test case. Thus, from each ``tag-sheet'' we can get an encoding of the test case layout as a String of any combination of \textit{``arrange''}, \textit{``act''}, \textit{``assert''}.

\textbf{\textit{Manual Tagging}} The taggers first review the test case name and the containing test class name. Good names usually help the taggers hold a quick grasp of the test case's intention. It is typical to see a test case name starting with the verb \textit{``test''}, followed by the function and scenario under test; and the test class name is usually \textit{``AlphaTest''}, where \textit{``Alpha''} is a higher summary of the tested functions. For example, test cases \textit{``testInvoker\_normal''} and \textit{``testInvoker\_fail''} are under test class \textit{``ClusterInvokerTest''}. Usually, the test case name reveals what is the function under test---pointing to the \textit{act}. For example, \textit{``testInvoker\_normal''} and \textit{``testInvoker\_fail''} should both \textit{act} \textit{``cluster.invoke()''}.

Next, the taggers dive deep into the internal logic of the test case. The taggers do not only review the lines of code within the test case, but also review the expanded code from methods defined in the test class and invoked by the test case. This is critical for gaining a flattened view of the full \textit{AAA} structure of a test case. For example, test case \textit{testReleaseDedicatedGuestVlanRange}~\cite{testReleaseDedicatedGuestVlanRange} from CloudStack only contains three lines of code by itself. But it calls the method \textit{runReleaseDedicatedGuestVlanRangePostiveTest}, which contains 9 expanded statements with all three ``A''s. 

In addition, the taggers often also have to carefully review the inside of the production functions called in the test case. This is especially important for tagging test cases that do not follow good naming conventions. For example, it is not rare to see a test case named ``testX'', where ``X'' is a number or an alphabet with no clue for the intention of the test case. For example, \textit{test2}~\cite{test2} is from Accumulo. Tagging the \textit{act} of such cases usually requires understanding of the production functions. In  \textit{test2}~\cite{test2}, the invocation (line 86) to \textit{MultiIterator.seek} is marked as \textit{act} since other called production methods are simple setups. 

In the tagging process, test cases emerge that they have statements for ``tearing down'' arranged objects, which should not be tagged as any of the ``A''s. This often appears when the test case uses static objects or attributes with global access by different test cases. Thus, we leave these statements instead of forcing them into any ``A'', which should not interfere with the understanding of the \textit{AAA} structure of a test case.

\textbf{\textit{Cross-validation.}} After tagging independently, the taggers compare and cross-validate their results. The cases with disagreements are brought into group discussions which involve two full-time developers from the industry. We use Cohen's kappa~\cite{mchugh2012interrater} to assess the agreement between the two taggers. This is assessed for the three ``A''s separately---with 0.97 on \textit{arrange}, 0.95 on \textit{act}, and 0.98 on \textit{assert}. The tagging of \textit{assert} is most straightforward---mostly relying on the JUnit Assert APIs. We also recognize some common syntax features for \textit{arrange}, such as invocations to \textit{setter}, \textit{constructor}, and \textit{mock} APIs. Tagging \textit{act} is most critical, relying on the understanding of the test case intention. The most frequently recurring disagreement scenario is that two taggers identify different \textit{act} on test cases with vague names, such as ``testX''. The other frequent disagreement is on \textit{assert}, when it is used for checking pre-condition with the arranged objects. One tagger treats it as \textit{assert}; while the other treats it as \textit{arrange}, which is our final tagging. 

\subsection{Step 2: Test Case Analysis under \textit{AAA} Context}
In this step, we first use regular expression matching to identify test cases that
follow the classic \textit{AAA} pattern. We then
manually examine the remaining test cases to identify cases that do not follow
the classic \textit{AAA} but should still be considered as \textit{AAA} due to special design. The remaining test cases
are considered violations of AAA. Next, we compare the complexity of
the test cases that follow \textit{AAA} and those not. Finally, we perform a thorough analysis of the test cases to identify recurring
anti-patterns that deviate from the \textit{AAA} structure and design flaws reside in the \textit{A} blocks of the \textit{AAA} test cases.

\textbf{\textit{Regex Matching.}} We take the encoding of each test case from the ``tag-sheet'' as input, and match it against a regular expression, namely ``[arrange]+[act]+[assert]+''. This expression implies that the encoding of a test case should be composed of at least one \textit{arrange}, followed by at least one \textit{act}, and followed by at least one \textit{assert}. 
If a test case matches the regular expression, it indicates that it follows the  \textit{classic AAA} pattern.

\textbf{\textit{Manual Inspection.}} For cases that do not match the regular expression, we revisit the source code to understand whether the \textit{AAA} pattern is truly violated. In the manual tagging process, we focus on analyzing inside the scope of each selected test case. Here, we expand our inspection of such a test case to the entire scope of the test class where the test case resides. There could be special design considerations. For example, a test case may access global \textit{arrange} or \textit{assert} shared across different test cases, encapsulated in the \textit{@Before} or \textit{@After} methods. We consider such cases as \textit{Special AAA} since the ``violation'' of \textit{AAA} pattern is superficial, but they essentially follow the \textit{AAA} pattern through special design. We summarize recurring design patterns that lead to \textit{Special AAA}. The \textit{AAA} test cases should contain both the classic \textit{AAA} and special \textit{AAA}. For the remaining cases, which truly violate the \textit{AAA} pattern due to improper design, we categorize them as \textit{Anti-AAA} cases.

\textbf{\textit{Comparison.}} We compare the complexity of \textit{AAA} cases and \textit{Anti-AAA} cases by the LOC and Cyclomatic complexity. We aim to understand if following the \textit{AAA} pattern leads to less complicated test cases. Furthermore, we also analyze the numbers of statements in test cases that are tagged as \textit{arrange}, \textit{act}, and \textit{assert}. This helps us to review the overall layout of the three \textit{``A''s}. For example, a unit test case following the \textit{AAA} should typically contain one \textit{act}, multiple \textit{arrange} and \textit{assert} statements. While in cases that violate \textit{AAA}, the layout could show bigger variations, especially in \textit{act}.
 
\textbf{\textit{Design Problem Identification.}}
For cases that violate the \textit{AAA} pattern, we analyze how and why the \textit{AAA} pattern is violated by reviewing both their encoding and source code. In particular, we reason about and make note of what is the drawback of violating the \textit{AAA} pattern in each case, and how could we address the structure. 

For cases that already follow the \textit{AAA} pattern, we focus on identifying potential design flaws within each \textit{A} block. In particular, we focus on the syntax of control flow, such as \textit{if-else}, \textit{for}, \textit{while}, and \textit{try-catch}. These imply that the execution logic of the test case relies on input conditions. They are focal points of complexity in the \textit{A} blocks, and likely where design flaws reside. Apparently, control-flow logic does not always leads to design flaws; instead, it could be necessary for certain testing purposes. Similarly, we record what problem we observe in each case, and reason about drawbacks and resolutions. 

The identification of recurring \textit{anti-AAA} patterns and design flaws is based on the ground theory~\cite{stol2016grounded}---constructed bottom-up from our notes. We will discuss these problems in detail in Section~\ref{sec:rq2}.




\subsection{Step 3: Refactoring and Feedback Collection}
In this step, we first implement the refactoring for different design problems we identified, and then send refactoring proposals to developers through issue tickets and pull requests for collecting their feedback. 

\textbf{\textit{Refactoring.}}
The first author reviews each problematic test case carefully again to attempt refactoring. For the \textit{AAA} cases with design flaws, the goal is to improve inside the \textit{A} blocks. For the \textit{Anti-AAA} cases, the objective is to restructure them to follow the classic \textit{AAA} pattern. He selects representative cases from each project and proposes tentative refactoring solutions to discuss with all authors (two are real-life developers) in weekly group meetings. The team discuss the solution, suggest improvements, and reason about the benefits of refactoring. This may trigger iterative discussions and improvements on a test case. Once approved by the research team, we proceed with the case to the next phase---preparing and submitting an issue ticket to the project. 

\textbf{\textit{Issue Tickets.}} With the research team's approval, the first author prepares and submits an \textit{Issue Ticket (IT)} describing the problem we found and the resolution we may offer. Initially, we started with a \textit{Pull Request (PR)} to Accumulo, which directly sends a refactoring solution for developers' review. One of the developers from Accumulo replied and suggested that we start with an \textit{Issue Ticket (IT)} to be less intrusive since \textit{PR} takes more resources to handle. If a \textit{IT} confirms that the problem is valuable and a \textit{PR} is welcome, we move on to a \textit{PR}. 

When creating an \textit{IT}, we describe 1) what problem is being identified; 2) how to fix the problem; and 3) what is the benefit of fixing the problem. Following is an example \textit{IT} we create for Dubbo. We found a test case \textit{assert} a precondition, and we suggest replacing \textit{assert} by \textit{assume}.

\begin{tcolorbox}[size=title, opacityfill=0.25,enhanced,breakable]
\textbf{Bring in the Junit Assume to \textit{testSubscription}}

Hi Dubbo Community, 

I noticed that test case \textit{testSubscription} is asserting the precondition (\textit{pList}) before the actual test target (\textit{multipleRegistry.subscribe()}) is executed. So when the test fails, it may be due to 2 reasons:  1) The functions related to the precondition (here is \textit{pList}) have bugs; and 2) The functions related to \textit{multipleRegistry.subscribe()} have bugs (what we want to check actually with this test). The test case should focus on the target function, not its preconditions. The precondition function should be checked by its own test cases and should not let this case fail. So here, I suggest replacing the Assert function and System.out.println in lines 147-148 with an Assume function (introduced after JUnit 4): \textit{Assume functions is a set of methods useful for stating assumptions about the conditions in which a test is meaningful. A failed assumption does not mean the code is broken, but that the test provides no useful information.}

\begin{itemize}
    \item Before the refactoring: 
\begin{lstlisting}[language = Java, frame = single , numbers=left, firstnumber = 145 , escapeinside={(*@}{@*)}]
String path = "/dubbo/" + SN;
List<String> pList = zkClient.getChildren(path);
Assertions.assertTrue(!pList.isEmpty());
System.out.println(pList.get(0));
\end{lstlisting}
    
    \item After the refactoring:
    
\begin{lstlisting}[language = Java, frame = single , numbers=left, firstnumber = 145 , escapeinside={(*@}{@*)}]
String path = "/dubbo/" + SN;
List<String> pList = zkClient.getChildren(path);
Assumptions.assumingThat(pList.get(0), is(ERROR_MSG));
\end{lstlisting}
\end{itemize}
After the replacement, when running this test case, developers can focus on the test target, and clearly identify the source of failure.


\end{tcolorbox}

\textbf{\textit{Pull Requests.}} We only move forward to \textit{PR} if we get an invitation from the developer to do so. For example, for the above \textit{IT}, we receive \textit{``Good idea. Would you please (submit) a PR?''}. In a similar \textit{IT} for Druid, a developer asked us for clarification of how \textit{assume} works, but he never returned to us, after we provided additional information. We did not proceed with the \textit{PR} since no explicit invitation is given. When preparing a \textit{PR}, we need to run through the CD/CI pipeline of the project to make sure it does not break a build. This may trigger additional correspondence with the developers and cause delays.

\section{Study Results}\label{sec:results}

\begin{figure}[h]
	\centering
    \includegraphics[width=\columnwidth]{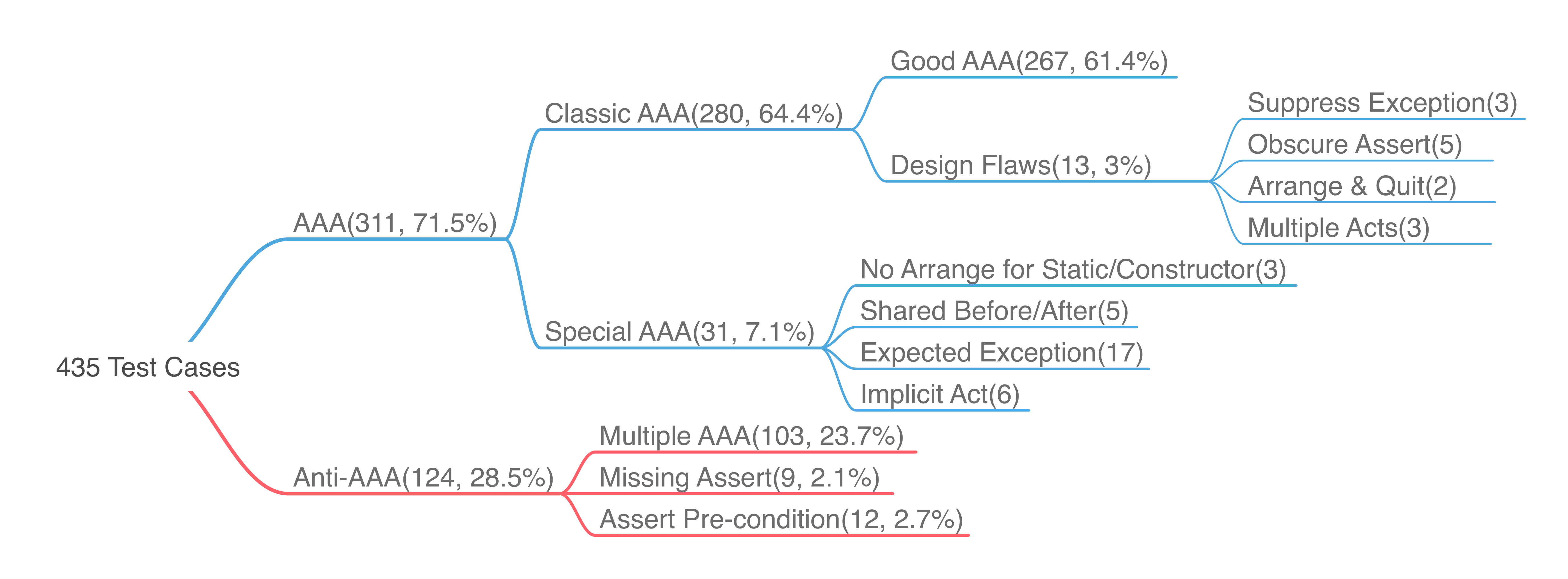}
	\caption{Test Cases Categorization}
	\label{fig:category}
\end{figure}

In manual tagging, we exclude 65 test cases from the initial dataset since they are not unit test cases, thus are out of the scope of this study. The \textit{AAA} pattern provides a uniform structure to unit test cases, but may not be appropriate to handle the complexity of other tests, such as integration tests. The excluded test cases are often integration tests that focus on testing a flow of functions, with names ending with an ``IT''. Some test cases wrap external commands and SQL queries, which are not unit test cases either. For the remaining 435 test cases, which---to our best understanding---are unit test cases, their categorization of whether they follow the \textit{AAA} pattern is illustrated in Figure~\ref{fig:category}. The details of this categorization will be discussed in answering the RQs below. The data is openly available on figshare\footnote{\url{https://figshare.com/s/b1d6b70e10837aaf3f17}}.

\subsection{RQ1: \textit{AAA} vs \textit{anti-AAA} }

Overall, 71.5\% test cases in our dataset follow the \textit{AAA} pattern. As shown in Figure~\ref{fig:category}, 280 (64\%) test cases follows the \emph{classic} \textit{AAA} pattern, which shows the \textit{arrange}, \textit{act}, and \textit{assert} layout. In addition, 31 test cases 
do not directly manifest the classic \textit{AAA} structure, but they
essentially follow the \textit{AAA} design due to special design considerations, which is explained in detail below:

\paragraph{\textbf{Special AAA}} We found four \textit{Special AAA} patterns due to special design considerations:


\textbf{No Arrange for Static/Constructor:} The target of the test case is a static function or a constructor, thus the test case does not need any arrangement. Following is such an example.
\begin{lstlisting}[language = Java, frame = single , numbers=right, firstnumber = 1 , escapeinside={(*@}{@*)}]
@Test
public void testStatic() {
    int a = SomeClass.aStaticMethod();
    assertEquals(1,a);}
\end{lstlisting}

\textbf{Shared Before/After}: The \textit{arrange} or \textit{assert} sections are encapsulated in methods with special annotation, such as @Before or @After. These methods, not explicitly called in any test case, execute before or after the execution of each test case due to the annotations supported by JUnit. The following code snippet illustrates such an example. 
\begin{lstlisting}[language = Java, frame = single , numbers=left, firstnumber = 1 , escapeinside={(*@}{@*)}]
@Before
public void setup(){
    data = new Data(src, dest);}
@After
public void verify(){
    assertNotNull(data.getValue());}
@Test
public void testConfigBig(){
    data.config("Big");}
\end{lstlisting}    
    
\textbf{Expected Exception}: A test case uses the \textit{expected} attribute of the \textit{@Test} annotation to declare that it expects an exception to be thrown. The following illustrates such an example. 
\begin{lstlisting}[language = Java, frame = single , numbers=left, firstnumber = 1 , escapeinside={(*@}{@*)}]
@Test(expected = ClientException.class)
public void testEmptyClientException() throws Exception {
    try(Client client =new Client("")){
        client.createProfile();}}
\end{lstlisting}

\textbf{Implicit Act}: The test case does not have an explicit \textit{act}; while the JUnit \textit{assert} function executes the \textit{act} function through dynamic binding. These cases are all associated with testing the \textit{equals} overridden by user-defined functions. The following is an example. 
\begin{lstlisting}[language = Java, frame = single , numbers=left, firstnumber = 1 , escapeinside={(*@}{@*)}]
@Test
public void testEquals() throws Exception {
    Client a = new Client("Bob");
    Client b = new Client("Bob");
    assertEquals(a, b);}
\end{lstlisting}

\begin{figure}[h]
\begin{subfigure}{0.45\columnwidth}
  \centering
  \includegraphics[width=\columnwidth]{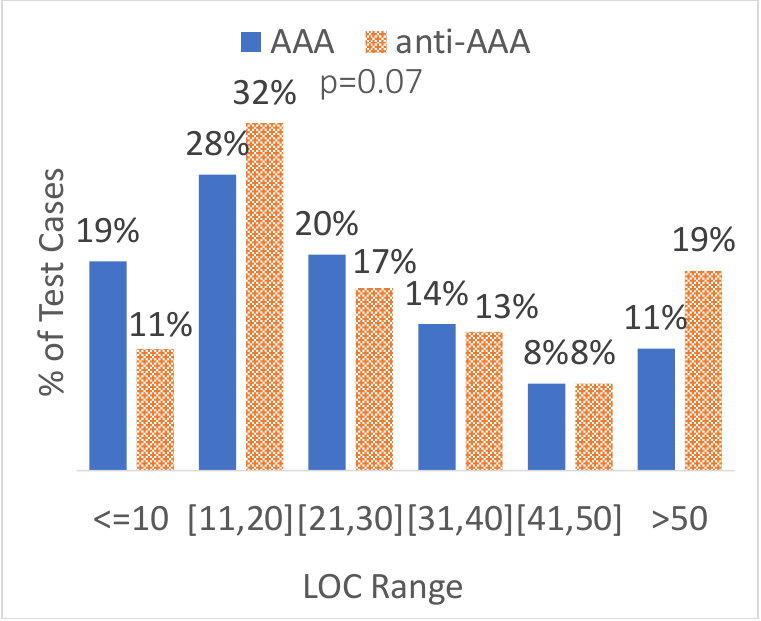}  
  \caption{\textit{LOC}}
  \label{fig:rq1-loc}
\end{subfigure}
\begin{subfigure}{0.45\columnwidth}
  \centering
  \includegraphics[width=\columnwidth]{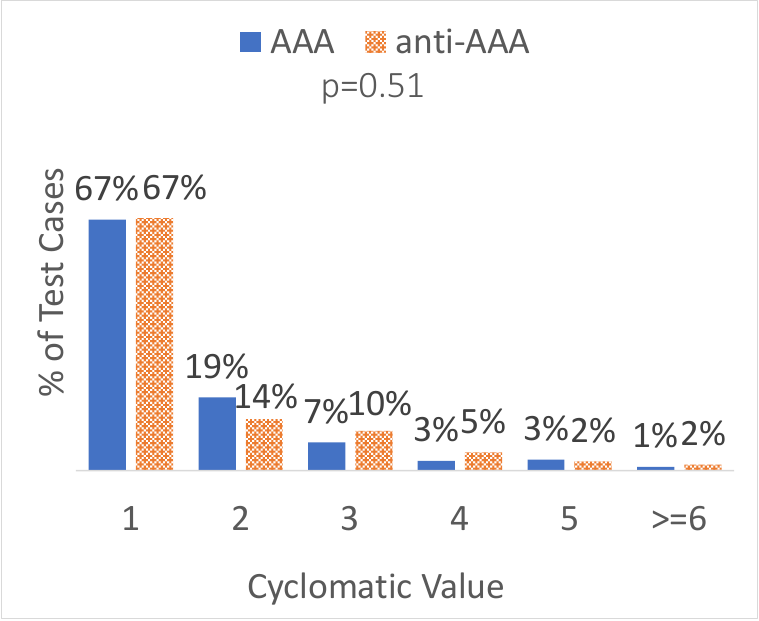}  
  \caption{\textit{Cyclomatic}}
  \label{fig:rq1-mc}
\end{subfigure}
\caption{Complexity of Test Cases }
\label{fig:rq1-complexity}
\end{figure}

\textit{\textbf{Comparison of AAA. vs. \textit{anti-AAA}.}} Figure~\ref{fig:rq1-complexity} compares the distribution of the LOC (Figure~\ref{fig:rq1-loc}) and Cyclomatic metric (Figure~\ref{fig:rq1-mc}) of \textit{AAA} vs. \textit{anti-AAA} test cases. The Figure~\ref{fig:rq1-loc} shows that the \textit{anti-AAA} cases tend to have slighter higher LOC than the \textit{AAA} cases, but the difference is not significant---p-value 0.07 ($>$0.05). Figure~\ref{fig:rq1-mc} shows that the Cyclomatic metrics for \textit{anti-AAA} and \textit{AAA} cases are undifferentiated---p-value 0.5. This indicates that the \textit{AAA} pattern does not have an obvious impact on the complexity of the test cases.


\begin{figure}[h]
\begin{subfigure}{0.32\columnwidth}
  \centering
  \includegraphics[width=\columnwidth]{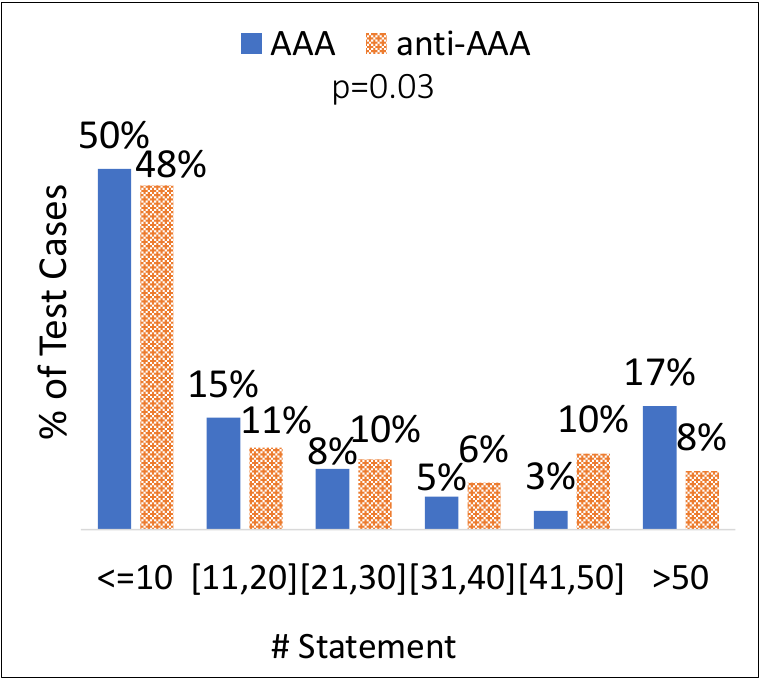}  
  \caption{\textit{Arrange}}
  \label{fig:rq1-arrangement}
\end{subfigure}
\begin{subfigure}{.32\columnwidth}
  \centering
  \includegraphics[width=\columnwidth]{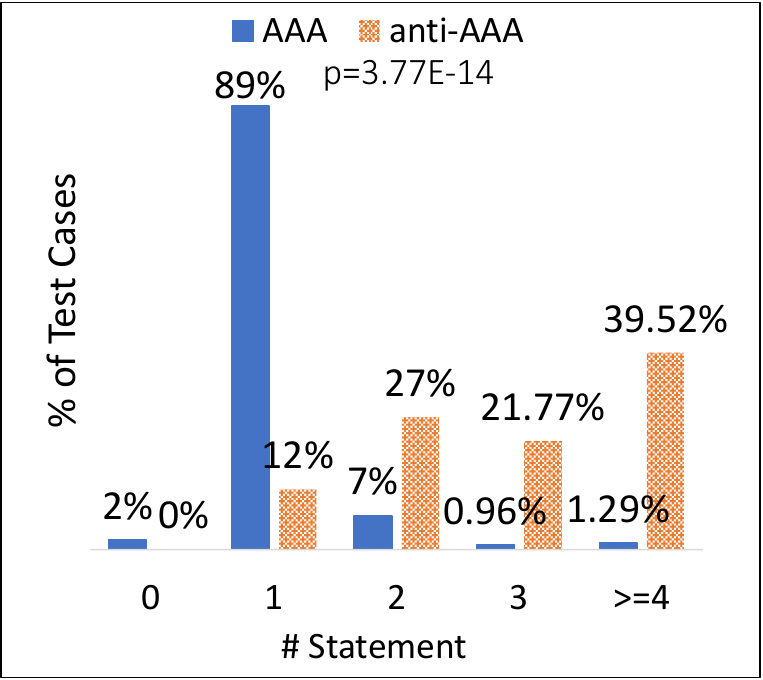}  
  \caption{\textit{Act}}
  \label{fig:rq1-action}
\end{subfigure}
\begin{subfigure}{.32\columnwidth}
  \centering
  \includegraphics[width=\columnwidth]{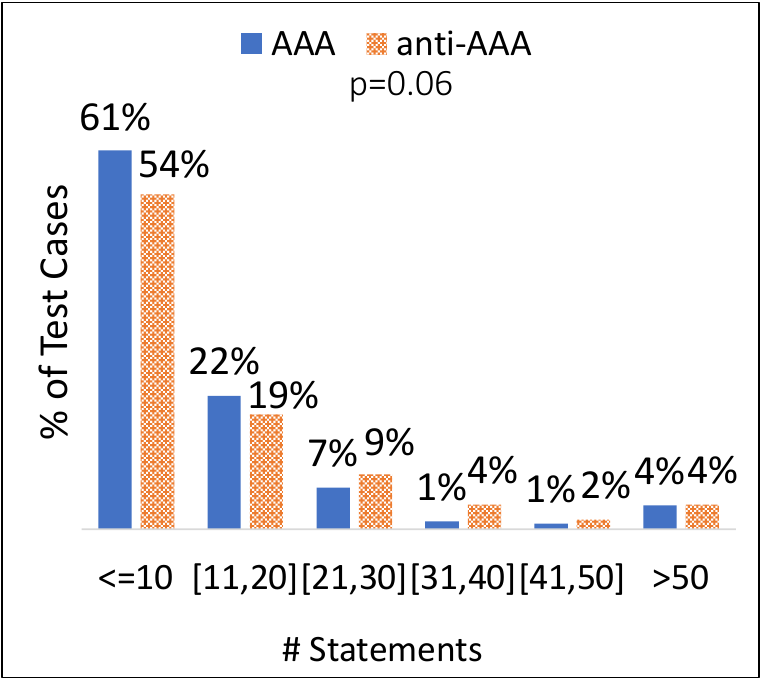}  
  \caption{\textit{Assert}}
  \label{fig:rq1-assertion}
\end{subfigure}
\caption{\# of Statements in Arrange, Act, and Assert}
\label{fig:rq1-statement}
\end{figure}

Figure~\ref{fig:rq1-statement} compares the \textit{anti-AAA} and \textit{AAA} cases in terms of their layout. That is the numbers of expanded statements as the \textit{arrange}, \textit{act}, and \textit{assert}, shown in Figure~\ref{fig:rq1-arrangement}, Figure~\ref{fig:rq1-action}, and Figure~\ref{fig:rq1-assertion} respectively. We can make three observations: 1) The \textit{AAA} cases seem to have a slightly higher number of \textit{arrange}---with a p-value of 0.03 ($<$0.5). The explanation is that developers tend to prepare more complicated \textit{arrange} for testing a target function when following the \textit{AAA} pattern. 2) The \textit{anti-AAA} cases obviously contain four times of \textit{act}s than \textit{AAA} cases---with p-value of 3.7E-14. And 3) there is no obvious difference in the number of \textit{assert} for the \textit{AAA} and \textit{anti-AAA} cases---p-value of 0.06 ($>$0.5). Therefore, the takeaway message is that the number of \textit{act} is the key difference between \textit{anti-AAA} and \textit{AAA} cases---whether the single responsibility 
principle~\cite{martin2003agile} is followed.




\begin{tcolorbox}[size=title, opacityfill=0.25,enhanced,breakable]
\textbf{RQ1 Summary:} Overall, 71.5\% test cases follow the \textit{AAA} structure, explicitly (64.4\%) or with some special design (7.1\%). Following the \textit{AAA} pattern does not have an obvious impact on the LOC or Cyclomatic complexity of test cases. The key difference is in the number of \textit{act}. Following the \textit{AAA} structure could be a way to facilitate the single responsibility principle in test case design---focusing only on one unit of function and one scenario. 
\end{tcolorbox}

\subsection{RQ2: Anti-AAA Patterns and Design Flaws}\label{sec:rq2}

\subsubsection{\textbf{Anti-AAA Patterns}}
 
 We summarized three recurring anti-patterns that deviate from the \textit{AAA} from the 51 cases shown in Figure~\ref{fig:category}. We explain each anti-pattern with a concrete example, its drawbacks, and corresponding refactoring resolution.

\textit{\textbf{Multiple AAA}}: The test case is composed of more than one \textit{AAA} blocks. For example, the following is an example with two blocks of \textit{AAA} combined in one test case. Line~\ref{b1-start} to line~\ref{b1-end}, the first block, intends to test \textit{getAllProperties} with \textit{PROP\_PREFIX}; while line~\ref{b2-start} to line~\ref{b2-end}, the second block, tests scenario \textit{SCAN\_PREFIX}. 

\begin{lstlisting}[language = Java, frame = single , numbers=right, firstnumber = 1 , escapeinside={(*@}{@*)}]
@Test//Multiple AAA
public void testGetByPrefix(){
    Config con = new Config();//arrange (*@\label{b1-start}@*)
    tc.set(PROP_PREFIX);//arrange
    var p = tc.getAllProperties();//act
    assertEquals("prop", p);//assert (*@\label{b1-end}@*)
    
    tc.set(SCAN_PREFIX);//arrange (*@\label{b2-start}@*)
    p = tc.getAllProperties();//act
    assertEquals("scan", p);}//assert(*@\label{b2-end}@*)
\end{lstlisting}

\textbf{Drawbacks:} First, the test case could get very large with multiple \textit{AAA} blocks combined, compromising the comprehension and maintenance of the test case. Second, there is more than one reason for the test case to fail, since each block of the \textit{AAA} could trigger a test failure. This leads to higher complexity in debugging. This anti-pattern is a violation of the single responsibility principle in software design~\cite{martin2003agile}.

\textbf{Refactoring:} Split it into multiple test cases, each containing one block of \textit{AAA} from the original case. As such, each test case follows the classic \textit{AAA} pattern, focuses on one test scenario, and should fail due to only one reason. If refactoring multiple AAA cases leads to multiple smaller but similar test cases that only vary in input parameters (i.e. code clone)~\cite{soares2022refactoring}, the developer could use the JUnit5 parameterized test feature to keep only one test case with annotated parameters to eliminate code clone. However, based on our observation, the refactored test cases from \textit{Multiple AAA} usually only share the same action, and will not lead to code clone.

\begin{lstlisting}[language = Java, frame = single , numbers=left, firstnumber = 1 , escapeinside={(*@}{@*)}]
@Test//Multiple AAA Refactored
public void testGetByPrefix_PROP(){
    Config con = new Config();//arrange 
    tc.set(PROP_PREFIX);//arrange
    var p = tc.getAllProperties();//act
    assertEquals("prop", p);}//assert 

@Test
public void testGetByPrefix_SCAN(){    
     Config con = new Config();//arrange
     tc.set(SCAN_PREFIX);//arrange 
     var p = tc.getAllProperties();//act
     assertEquals("scan", p);}//assert 
\end{lstlisting}

\textit{\textbf{Missing Assert}}: The test case does not contain any JUnit \textit{assert} function or does not specify any expected behavior (e.g. using the \textit{expected} attribute, as we described earlier). In other words, the test case will never raise a failure regardless of the correctness of the function under the test. The test case may use the print function, which delegates the inspection of results to manual effort. Following is an example.  
 \begin{lstlisting}[language = Java, frame = single , numbers=left, firstnumber = 1 , escapeinside={(*@}{@*)}]
@Test//Missing Assert
public void testDataGenerator(){
    Data d = new Data();//arrange
    d.generate();//act
    printData(d.getData());}
private void printData(var input){
    for(String d:input){
        System.out.println(d);}}
\end{lstlisting}

\textbf{Drawbacks:} If a test case never fails, it forfeits its purposes for capturing defects in the function under test. With the print/log method for manual checking, the cost is prohibitive, especially in the CD/CI environment of modern software development.

\textbf{Refactoring:} Add \textit{assert} function. For the \textit{print} method, it is often used when the result is long. The expected value could be stored in external resources, and loaded into the test case for assertion. The following illustrates the refactoring of the above example:

 \begin{lstlisting}[language = Java, frame = single , numbers=left, firstnumber = 1 , escapeinside={(*@}{@*)}]
@Test//Missing Assert Refactored
public void testDataGenerator(){
    Data d = new Data();//arrange
    d.generate();//act
    Vector<String> exp = load("gen.dat");
    for(String d:input){
        assertEquals(exp,d);}}
\end{lstlisting}

\textit{\textbf{Assert Pre-condition}}: The test case asserts certain pre-condition of the \textit{arranged} objects before \textit{acting} the function under test. As shown below, \textit{assertNull} makes sure that the \textit{snapshot} is acquired (i.e. not null) from the database. 

\begin{lstlisting}[language = Java, frame = single , numbers=right, firstnumber = 1 , escapeinside={(*@}{@*)}]
@Test//Assert Pre-condition
public void testPoll(){
   Snapshot s = sqlMng.createSnapshot();
   assertNotNull(s);
   String v = s.poll();
   assertEquals("8/22/2022",v);}
\end{lstlisting}

\textbf{Drawback:} The test case could fail due to two reasons: 1) the pre-condition is not met; 2) the function under test contains errors. This will add complexity to debugging. Also, the actual \textit{act} is not executed if the case fails due to the pre-condition.  

\textbf{Refactoring:} Replace the \textit{assert} pre-condition by the Junit \textit{assume}. The \textit{assume} is a set of methods introduced since JUnit 4 for stating assumptions about the conditions in which a test is meaningful. A failed \textit{assume} does not mean the code is broken, but that the test provides less useful information. 

\begin{lstlisting}[language = Java, frame = single , numbers=right, firstnumber = 1 , escapeinside={(*@}{@*)}]
@Test//Assert Pre-condition Refactored
public void testPoll(){
   Snapshot s = sqlMng.createSnapshot();
   assumeNotNull(s);
   String v = s.poll();
   assertEquals("8/22/2022",v);}
\end{lstlisting}

\subsubsection{\textbf{Design Flaws in \textit{AAA} Test Cases}}
Following the \textit{AAA} pattern does not imply that the test case is perfect. Here we present four types of design flaws we observed from the \textit{AAA} cases:

\textit{\textbf{Obscure Assert:}} The \textit{assert} block contains unnecessary control flow that obscures the logic of what is asserted. As shown below, the \textit{for} loop plus the \textit{if-else} block asserts that the elements in a collection satisfy certain conditions.
    \begin{lstlisting}[language = Java, frame = single , numbers=right, firstnumber = 1 , escapeinside={(*@}{@*)}]
@Test//Obscure Assert
public void testCluster(){
    ...
    Boolean foundValid = false;
    for(int cluster:clusterList){
        if(cluster != 1){fail("Err");}
        else{ foundValid = true;}}
    assertTrue(foundValid);}
\end{lstlisting}

\textbf{Drawback:} As the name suggests, this design flaw adds unnecessary complexity to the assert logic, obscures the intention of assert, and adds difficulty to comprehension and maintenance.

\textbf{Refactoring:} Eliminate unnecessary control flow to simply the logic of \textit{assert}. The above example can be simplified into one single \textit{assert} statement by using the \textit{hamcrest} API shown below. Note that not every such case requires the use of \textit{hemcrest}. 
\begin{lstlisting}[language = Java, frame = single , numbers=right, firstnumber = 1 , escapeinside={(*@}{@*)}]
import org.hamcrest.Matchers.*;
@Test//Obscure Assert Refactored
public void testCluster(){
    ...    
    assertThat(clusterList, everyItem(equalTo(1));}
\end{lstlisting}

\textit{\textbf{Arrange\&Quit:}} The test case \textit{returns} silently if the \textit{arranged} object does not meet certain condition, which is the counter-part of \textit{Assert Precondition}. See below, the test case \textit{returns} when \textit{thr} is \textit{null}.
\begin{lstlisting}[language = Java, frame = single , numbers=left, firstnumber = 1 , escapeinside={(*@}{@*)}]
@Test//Arrange&Quit
public void testSetException(){
    Throwable thr = buildXExp();
    if (thr == null) {return;}
    App app = new App().setExp(thr);
    assert.Equals(0, app.getMsg());}
\end{lstlisting} 

\textbf{Drawbacks:} First, the \textit{if-return} makes the logic the test case implicit. Second, the test case will quit silently without any hint regarding if the test case executed successfully or not. If not, why.

\textbf{Refactoring:} Replace the \textit{if-return} block by an \textit{assume} for the precondition, shown below:
\begin{lstlisting}[language = Java, frame = single , numbers=left, firstnumber = 1 , escapeinside={(*@}{@*)}]
@Test//Arrange&Quit Refactored
public void testSetException(){
    Throwable thr = buildXExp();
    assumeNotNull(thr);
    App app = new App().setExp(thr);
    assert.Equals(0, app.getMsg());}
\end{lstlisting} 

\textit{\textbf{Multiple Acts:}} The test case \textit{acts} more than one functions of a class. Following is an example from CloudStack. The test case name \textit{testCreateAndInfoWithOptions} suggests that it aims at testing both the creation and the getting of info. 
\begin{lstlisting}[language = Java, frame = single , numbers=left, firstnumber = 1 , escapeinside={(*@}{@*)}]
@Test//Multiple Acts
public void testCreateAndInfo(){
    ...
    qemu.create(file); (*@\label{create}@*)
    Map info = qemu.info(file); (*@\label{info}@*)
    ...}
\end{lstlisting}

\textbf{Drawbacks:} If the test case fails, it is difficult to tell which function leads to the failure. In addition, each individual function is usually not adequately asserted, since the \textit{assert} focuses on the final output, but overlooks the intermediate output.

\textbf{Refactoring:} Break the test case into separate ones, each focusing on one \textit{act} and add separate \textit{assert}:
\begin{lstlisting}[language = Java, frame = single , numbers=left, firstnumber = 1 , escapeinside={(*@}{@*)}]
@Test//Multiple Acts Refactored
public void testCreate(){
    ...
    qemu.create(file); (*@\label{create}@*)
    assertTrue(file.exist())
    ...}
@Test
public void testInfo(){
    ...
    qemu.create(file); (*@\label{create}@*)
    Map info = qemu.info(file); (*@\label{info}@*)
    assertEquals(SIZE, info.size());
    ...}
\end{lstlisting}   

On a particular note, multiple \textit{act} does \textit{not} always lead to design flaws. It's legitimate when the test case aims to test the repeated execution of a function. Or the target function requires calling a set of related methods, which could be an indication of poor production API design. 



\textit{\textbf{Suppressed Exception:}} The test case uses the \textit{try-catch} block to suppresses an \textit{Exception} that should be thrown and raise a failure. As shown below, the \textit{try-catch} block catches the \textit{Exception} and prints the stack trace.

\begin{lstlisting}[language = Java, frame = single , numbers=right, firstnumber = 1 , escapeinside={(*@}{@*)}]
@Test//Suppressed Exception
public void testHttpclient() {
    ...
    try { client.execute(); } //act 
    catch (final ClientException e) {
        e.printStackTrace();}
    ...}
\end{lstlisting}

\textbf{Drawbacks:} This suppresses the \textit{Exception} and will not raise a failure. It hides the error from developers.

\textbf{Refactoring:} Remove the \textit{catch} and keep the \textit{try}. Add \textit{throws} for the Exception that was caught initially, shown below:

\begin{lstlisting}[language = Java, frame = single , numbers=right, firstnumber = 1 , escapeinside={(*@}{@*)}]
@Test//Suppressed Exception Refactored
public void testHttpclient() throws ClientProtocolException {
    ...
    client.execute(); //act
    ...}
\end{lstlisting}


In this study, we follow the definition of \textit{test refactoring} by van Deursen et. al. ~\cite{Deursen2001RefactoringTC} ``do not add or remove test cases" and improve its quality. Of particular note, different from refactoring the production code, refactoring certain test cases are intended to change the behavior to make the test case more powerful. More specifically, \textit{Missing Assert} is refactored by adding an assertion to report failures; \textit{Assert precondition} is refactored that the precondition should not fail the test case; \textit{Arrange\&Quit} is refactored to use assume such that the action will get executed anyways; \textit{Suppressed Exception} is refactored to expose the exception. All the other refactoring types preserve the behavior of the original test case.

In addition, some of the refactoring types are highly generalizable and can be even automatable, such as Assert Pre-condition, Arrange\&Quit, and suppressed Exception. Others are less likely to be fully automated due to the case-by-case complexity, such as multiple AAA (where and how to break) and Missing Assert (what is the assert condition). This calls for more future studies, for which this empirical study provides insights.

\begin{tcolorbox}[size=title, opacityfill=0.25]
\textbf{RQ2 Summary:} We observed three recurring \textit{Anti-AAA} patterns---\textit{Multiple AAA}, \textit{Missing Assert}, and \textit{Assert Pre-condition}, as well as four design flaw types that reside inside of the \textit{A} blocks---\textit{Suppress Exception}, \textit{Obscure Assert}, \textit{Arrange\&Quit}, and \textit{Multiple Act}. Each problem type has its own drawbacks and corresponding refactoring resolutions.
\end{tcolorbox}

\subsection{RQ3: Developers' Feedback}
\subsubsection{\textbf{Anti-AAA Patterns}}

The first half of Table~\ref{tbl:issues} lists the 11 proposals we submit. They are selected to cover three anti-patterns in all projects, except \textit{Assert Precondition} in Accumulo, which does not exist. The table shows the project, the test case, the problem type, the turn-around time, and the status of the \textit{IT} and \textit{PR} we submit. 


As we can see that we have a 100\% response rate, with most responding in a few days after it is sent out, except the proposal for test case \textit{testAutoRenewalDisabled}(row 6) from CloudStack, which takes 4 months. This indicates that the proposals are overall of interest to the projects. 

We received a positive response for 7 (64\%) \textit{ITs}. Especially, in the 3 \textit{ITs} for Dubbo, the developers proactively created respective \textit{PRs}, all promptly merged to the project. For the 4 \textit{ITs} on row 4 to row 7, the developers invited us to submit a \textit{PR} accordingly, such as \textit{``changes sound good to me. It will be great if you can create a PR.'', ``Could you please propose a PR?'', etc.} 

The \textit{PR} for \textit{testIsTaskCurrent}(row 4) is pending due to the CI/CD pipeline failure, which also happened to other \textit{PRs} sent at the time, thus it is not caused by our change. The another PR(row 16) has been pending review for a few months. We believe this relates to the priority of different \textit{PRs}. Usually, general improvement \textit{PRs} have a lower priority compared to bugs or features. However, the \textit{PR} invitation in our \textit{IT} response is already an indication of interest in fixing these anti-patterns.

The \textit{IT} (row 8) for \textit{testPollOnDemand} is \textit{``Ask for Info.''}, since the developer asked for more information about the \textit{assume}. 
However, the developer has not gotten back to us after we provided more clarification a few months ago.

Three \textit{ITs} are rejected as shown from row 9 to row 11 in Table~\ref{tbl:issues}. They happen to cover all three anti-pattern types, indicating that developers may not have a strong preference for a particular anti-pattern type. The \textit{IT} for \textit{testCreateSuccess} (row 9) proposes to add \textit{assert}. A developer responded that ``\textit{I think you can leave this. if the creation is not successful it would have thrown an exception. On the other hand, this is one that is tested a lot of times implicitly as well.}''. Thus, we did not follow up with a \textit{PR}. 

The proposal for the \textit{Multiple AAA} in test case \textit{testGetByPrefix} (row 10) in Accumulo was our first attempt. Without prior experience, we directly sent a PR. This is where we received the suggestion to first send an \textit{IT} to be less intrusive and we followed this suggestion afterward. 
Although the developer rejected our refactoring proposal, he/she shared very insightful feedback with us, detailing the considerations regarding why the refactoring is not favored: 1) The developer agreed that each new test is simpler and more granular on its own, but they do not need every test case to be that granular. They prefer to keep the original test case with a bunch of trivial cases around a single method. The developer is concerned that if they were to break test cases into this level of granularity, the code would become indiscriminately large and unwieldy. And, 2) the developer pointed out that this test never fails. Thus, the benefit of refactoring it does not justify the investment in reviewing, verifying, and approving this change. 

The same developer also saw the \textit{PR} of \textit{testSasl} (row 11) for replacing \textit{Assert Precondition} by \textit{Assume}. He expressed that this type of general improvement does not match the needs of any coherent plan of Accumulo. However, two other developers from Accumulo responded positively to our proposal to refactor the \textit{Obscure Assert}. And one of them even encouraged us to also submit improvements to similar test cases that he is aware of having similar issues. This indicates that different developers may have different takes on general improvements to test cases.

\begin{table*}[h!]
\caption{Summary of Refactoring Tickets}\label{tbl:issues}
\begin{adjustbox}{width=\textwidth,center}
\centering
\begin{threeparttable}
\begin{tabular}{|l|l|l|l|l|l|l|}
 \hline
 ID
 &Project 
 &Test Case 
 &Issue Type 
 &Turn Around 
 &IT Status 
 &PR Status 
 \\ \hline

1&Dubbo & testAll~\cite{testAll} & Multiple AAA & 2 Days & Internal PR  & Merged \\ \hline

\multirow{2}{*}{2}&\multirow{2}{*}{Dubbo} & \multirow{2}{*}{testClear~\cite{testClear}} & Missing assertion & \multirow{2}{*}{10 Days} & \multirow{2}{*}{Internal PR} &\multirow{2}{*}{Merged} \\ 
&&& \& Assert Precond.  & & & \\ \hline

3&Dubbo & testSubscription~\cite{testSubscription} & Assert Precond.& 2 Days& Internal PR & Merged \\ \hline

4&Druid & testIsTaskCurrent~\cite{testIsTaskCurrent} & Multiple AAA & 3 Days & PR Invitation & Submitted \\ \hline
 
5&Druid & testNormal~\cite{testNormal} & Missing Assert  & 2 Days & PR Invitation & Merged \\ \hline

6&CloudStack & testAutoRenewalDisabled~\cite{testAutoRenewalDisabled} & Assert Precond. & 4 Mon. & PR Invitation & Merged\\ \hline
 
7&CloudStack &  testCRUDacl~\cite{testCRUDacl} & Multiple AAA  & 1 Days & PR Invitation & Merged \\ \hline

8&Druid & testPollOnDemand~\cite{testPollOnDemand} & Assert Precond. & 4 Days & Ask for Info. & - \\ \hline
 
9&CloudStack & testCreateSuccess~\cite{testCreateSuccess}& Missing Assert  & 1 Day & Reject & - \\ \hline

10&Accumulo& testGetByPrefix~\cite{testGetByPrefix} & Multiple AAA & 1 Day & -* & Reject \\ \hline

11&Accumulo & testSasl~\cite{testSasl} & Assert Precond. & 1 Day & -* & Reject \\ \hline 
\multicolumn{7}{c}{}\\ \hline
12&CloudStack & testHttpclient~\cite{testHttpclient}& Suppress Exception & 1 Day & PR Invitation & Merged \\ \hline
13&Accumulo & verifyExceptionCallingStartWhenRunning~\cite{verifyExceptionCallingStartWhenRunning} & Obscure Assert & 1 Day & PR Invitation & Merged \\ \hline 
14&CloudStack & searchForNonExistingLoadBalancer~\cite{searchForNonExistingLoadBalancer} & Obscure Assert & 1 Day & PR Invitation & Merged\\ \hline
15&Druid & testReadParquetDecimali32~\cite{testReadParquetDecimali32} & Arrange \&Quit & 1 Day & PR Invitation & Merged\\ \hline

16&CloudStack & testCreateAndInfo~\cite{testCreateAndInfo}&Eager Test & 1 Day& PR Invitation & Submitted\\ \hline

17&Dubbo & testSetExceptionWithEmptyStackTraceException~\cite{testSetExceptionWithEmptyStackTraceException} & Arrange \&Quit & 13 Days & PR Invitation & Merged \\ \hline

18&CloudStack & checkStrictModeWithCurrentAccountVmsPresent~\cite{checkStrictModeWithCurrentAccountVmsPresent} & Obscure Assert & 3 Days & No Response* & Merged\\ \hline

\end{tabular}
 \begin{tablenotes}
      \small
      \item -*These were the first two we sent out directly as PRs. A developer suggests always starting with an IT. We followed this instruction.
      \item *We sent the PR without an IT response, since a developer commented in a prior \textit{IT} that we could directly send 
\end{tablenotes}
\end{threeparttable}
\end{adjustbox}
\end{table*}

\subsubsection{\textbf{Design Flaws in AAA}} The bottom half of Table~\ref{tbl:issues} lists the 7 proposals we submit. Note that we only found \textit{Suppress Exception} and \textit{Eager Test} from CloudStack. But the other two types, \textit{Obscure Assert} and \textit{Arrange\&Quit} are found and reported across different projects. 

We received 7 (100\%) responses. In 6 \textit{ITs}, developers invited us to submit follow-up \textit{PRs}. For example, ``\textit{Seems reasonable - can you submit a PR against the main branch?}'', ``\textit{You can open a PR to improve this.}'', ``\textit{please go ahead and create your PR, looking forward to it.}''. We did not receive any response for the \text{IT18} (row 18). But a developer commented in a prior \textit{IT} that we could send \textit{PR} directly in the future. Thus, we send the \textit{PR} without a response to the \textit{IT}, which is promptly merged.

\begin{tcolorbox}[size=title, opacityfill=0.25, enhanced, breakable]
\textbf{RQ3 Summary:} the 18 proposals received 100\% response rate. 78\% responses are positive---we are invited to submit a \textit{PR} or a \textit{PR} is merged. This indicates that real-life developers care about the design of test cases, and they are interested in fixing the problems we identified. Rejections also point to valuable lessons---return-on-investment is a key concern, which could consider the change- and failure-proneness of test cases, and granularity of change. 
\end{tcolorbox}

\section{Threat to Validity and Limitations}
\label{sec:limit}
We acknowledge that our study does not comprehensively capture all possible design problems in unit test cases. We cannot guarantee that the four design flaws in \textit{AAA} cases have captured all possible problems in the \textit{A} blocks. In addition, we cannot guarantee that there are no other \textit{Anti-AAA} patterns. 

We acknowledge that the developers' feedback in our study cannot represent the opinions of other developers. In particular, we found that different developers, even from the same project, may have different perspectives on general improvements to test cases. Thus, there is intrinsic subjectivity from the developers who engage with our proposals. However, it is reasonable to claim that developers do generally care about the design of test cases. 

Finally, the study results may be subject to individual bias and experience in tagging \textit{AAA}, categorization, design issue identification, and refactoring. This is a threat to validity that is faced by any empirical study with manual effort and human intelligence. However, we made our best effort to mitigate this by engaging in team effort and weekly discussions. Also, we put significant effort into collecting feedback from developers. Therefore, it is reasonable to claim that our findings are valid and reflect real problems.
\section{Related Work}\label{sec:related}
In this section, we compare the design problems found in our study with test smells in the literature. According to Fowler, a code smell is a surface indication that usually corresponds to a deeper problem in the system~\cite{beck1999bad}. Test smells is a special group of code smells that appear in test code. Numerous prior studies have examined different types of test smells and their impacts~\cite{bavota2012empirical, tufano2016empirical, bavota2015test, spadini2018relation, peruma2019distribution, virginio2019influence, garousi2018we, qusef2019exploratory, de2019assessing, spinola2019understanding, garousi2018smells, athanasiou2014test, junior2020survey, peruma2020tsdetect}. Garousi and Küçük~\cite{garousi2018smells} present the largest catalogue of test smells derived from 166 sources. Most recently, Kim et al. investigated the evolution and maintenance of a comprehensive set of 18 test smells~\cite{kim2021secret}. 

Our study distinguishes itself for focusing on root-cause revealing design problems by leveraging the holistic \textit{AAA} context in a test case. In comparison, code smells suffer from staying at the surface of problem indications, as described by Fowler~\cite{beck1999bad}. For example, from the perspective of code smells, \textit{Duplicate Assert} points to test cases that \textit{assert} the same condition multiple times. One would naturally think that the solution is to remove the duplication, which could be misleading and erroneous. It is possible that the underlying root cause is our \textit{Multiple AAA}. Different test scenarios of a function repeats the same \textit{assert} multiple times---though there is repetition, it is necessary due to the logic of multiple test scenarios in one case. Without awareness of the \textit{AAA} context, one would easily fall into the pitfall of fixing the symptom but not the root cause. Each anti-pattern we identified is reasoned based on the AAA structure, which is not considered in smells. Following, we make a detailed  comparison of issues in our study with related test smells:


\begin{itemize}[noitemsep,topsep=0pt]
    \item Our \textit{Assert Precondition (AP)} and \textit{Arrange\&Quit (AQ)} may sound relevant to \textit{Rotten Green (RG)}~\cite{martinez2020rtj}. But a \textit{RG} test NEVER executes its Assert; while our \textit{AP} and \textit{AQ} do not even execute the Action under certain pre-condition. Of a particular note, ~\cite{de2008parameterized} advocated the usage of "assume", but we are the first to report \textit{Assert Precondition (AP)}, where "assert" is used inappropriately 
   instead of ''assume". 
    
    \item \textit{Obscure Assert (OA)} sounds similar to but is actually different from \textit{Assertion Roulette (AR)}~\cite{Deursen2001RefactoringTC}, \textit{Redundant Assertion (RA)}~\cite{peruma2019distribution}, \textit{Nested Conditional (NC)}~\cite{neukirchen2008approach}, and \textit{Duplicate Assert (DA)}~\cite{peruma2019distribution}. \textit{AR} contains multiple unexplained \textit{assert}; \textit{RA} \textit{asserts} a condition that is always true or always false; \textit{NC} contains nested conditional expression; and \textit{DA} \textit{asserts} the same condition multiple times. In comparison, our \textit{OA} points to the problem where it is obscure what is being asserted, e.g. assert in a loop that could be simplified.
    
    \item Our \textit{Multiple AAA} could be the underlying root cause of \textit{Assertion Roulette (AR)}---multiple unexplained \textit{assert}~\cite{Deursen2001RefactoringTC} and \textit{Duplicate Assert (DA)}---asserting the same condition multiple times~\cite{peruma2019distribution} to guide appropriate refactoring resolution. 
    
    \item Our \textit{Suppress Exception} takes a different perspective from the \textit{Exception Catch/Throw}~\cite{peruma2019distribution}. The literature generally considers using \textit{Exception Catch/Throw} as being problematic. However, our \textit{Suppress Exception} suggests that we should \textit{Throw} rather than \textit{Catch} an exception so as to expose it through test failure. 
    
    \item Our \textit{Missing Assert (MA)} is the same as \textit{Unknown Test}~\cite{peruma2019distribution}. And \textit{MA} may also suffer from \textit{Print Statement}~\cite{peruma2019distribution}. According to the literature, \textit{Print Statement} in test cases is generally problematic. We observe that some \textit{MA} cases use print for manual verification. But having \textit{Print Statement}  does not always imply \textit{MA}.
    
    \item \textit{Multiple Act} is the same as \textit{Eager Test}\cite{Deursen2001RefactoringTC}. 
    
\end{itemize}

\begin{tcolorbox}[size=title, opacityfill=0.25, enhanced, breakable]
\textbf{Take-away Message:} Although \textit{Missing Assert} and \textit{Multiple Act} overlap with two test smells, our study contributes four new problems---\textit{Multiple AAA}, \textit{Assert Precondition}, \textit{Arrange\&Quit}, and \textit{Obscure Assert}. In particular, our \textit{Multiple AAA} reveals underlying design root causes to several test smells, which is critical for proper refactoring. Finally, our \textit{Suppress Exception} stands at a different perspective from \textit{Exception Catch/Throw}. 
\end{tcolorbox}
\section{Conclusion}\label{sec:conclusion}
We conducted an empirical study of 435 unit test cases randomly selected from four open source projects. The objective was to understand whether \textit{AAA} is often followed in practice, identify design problems under the context of \textit{AAA} that merit refactoring, and collect developers' feedback on the refactoring. It turned out that 71.5\% of test cases indeed follow the \textit{AAA} structure---indicating that \textit{AAA} is well practiced. We discovered three recurring anti-patterns in test cases that deviate from the \textit{AAA} structure, and four types of design flaws that reside inside of the \textit{A} blocks, which merit from corresponding refactoring resolutions. The 18 representative proposals for fixing these problems are well-received by developers---with a 100\% response rate and 78\% responses favoring the refactoring. From the rejections, we learned that developers are concerned about the return-on-investment of such refactoring, considering the change-proneness, failure-proneness, and granularity of change. 

\IEEEpeerreviewmaketitle
\bibliographystyle{IEEEtran}
\bibliography{test}

\end{document}